\documentclass[12pt]{article}
\pdfoutput=1
\usepackage{pstricks}
\usepackage{color}
\usepackage{amssymb,amsmath,bm,bbm}
\usepackage{epsf}
\usepackage{epsfig}
\usepackage{afterpage}
\usepackage{longtable}
\usepackage{appendix}
\usepackage{cite}
\usepackage{latexsym,mathrsfs,dsfont}
\usepackage{graphics}
\usepackage{url}
\usepackage{paralist}
\usepackage{bbold}
\usepackage[colorlinks=true, urlcolor=blue]{hyperref}
\newcommand{\nn}{\nonumber}

\newcommand{\spur}[1]{\not\! #1 \,}
\newcommand{\be}{\begin{equation}}
\newcommand{\ee}{\end{equation}}
\newcommand{\bea}{\begin{eqnarray}}
\newcommand{\eea}{\end{eqnarray}}

\newcommand{\dd}{\displaystyle}
\begin{document}

\begin{flushright}
    {BARI-TH/25-775}
\end{flushright}

\medskip

\begin{center}
{\Large\bf
  \boldmath{
Two-lepton tales:\\ 
  \vspace*{0.3cm}
  Dalitz decays of  heavy quarkonia}}
\\[0.8 cm]
{\large P.~Colangelo$^{a}$, F.~De~Fazio$^{a}$ and R.~Pinto$^{b}$
 \\[0.5 cm]}
{\small
$^a$
Istituto Nazionale di Fisica Nucleare, Sezione di Bari, Via Orabona 4,
70126 Bari, Italy\\
$^b$ Dipartimento Interateneo di Fisica "Michelangelo Merlin", Universit\`a degli Studi di Bari, \\ via Orabona 4, 70126 Bari, Italy}
\end{center}

\vskip0.5cm


\begin{abstract}
\noindent
We study the Dalitz decays of heavy quarkonia, which result from the internal virtual photon  conversion   into an $\ell^+ \ell^-$ lepton pair. Heavy-quark symmetries allow us to establish systematic relations between transitions of different quarkonium states,  and to precisely determine the  branching fractions for several  charmonium and bottomonium decay modes. 
For charmonium,  existing data on $\chi_{cJ}(1P)\to J/\psi \ell^+ \ell^-$ and $\psi(2S)\to \chi_{cJ}(1P) \ell^+ \ell^-$
 enable us to determine the parameters of the  transition form factors  and to predict the rates of  yet-unobserved modes. The  Dalitz transitions of
 $\chi_{c1}(3872)$ are important, as  they can help assessing the structure of this meson.
For bottomonium,  recent LHCb measurements allow us  to predict the branching fractions  of $\chi_{bJ}(nP)\to \Upsilon(1S)\ell^+ \ell^-$ and $h_b(nP)\to \eta_b(1S) \ell^+ \ell^-$ ($n=1,\,2)$. We  also investigate the sensitivity of heavy quarkonia Dalitz modes to the contribution of a new light vector mediator,  such as the  putative  $X(17)$.
\end{abstract}

\thispagestyle{empty}
\newpage
\section{Introduction}

 The Dalitz decays of mesons  \cite{Dalitz:1951aj},  electromagnetic process where a virtual photon converts into a  charged lepton pair  (known as the internal photon conversion  mechanism), are important   probes of   hadron structure. Inspection of the Review of Particle Properties (PDG) indicates that several Dalitz modes have been  observed  and studied in detail for  light mesons \cite{ParticleDataGroup:2024cfk,Landsberg:1985gaz,Faessler:1999de}. 
Conversely, only  a few measurements are available in the charmonium sector  and none in the bottomonium sector. Nevertheless, there is a substantial  interest in the Dalitz modes of heavy quarkonia.  These processes are valuable to study QCD properties, specifically the consequences of spin symmetry emerging in the heavy quark limit \cite{Isgur:1991wq,Neubert:1993mb}.
 Moreover,  they can aid in assessing the structure of controversial states, with $\chi_{c1}(3872)$ meson (formerly $X(3872)$) being the prime example. An additional motivation  is their potential sensitivity to non-Standard Model contributions, as the contribution of light vector  mediators very feebly  coupled to Standard Model (SM)  particles,   which are generally referred to as dark photon $(\gamma^\prime)$ or dark $Z$. The hypotetical $X(17)$,  whose existence  has been invoked in connection with the ATOMKI anomaly \cite{Krasznahorkay:2015iga}, is one of such examples.

For  point-like particles,  the  $M^\prime  \to M \ell^+ \ell^-$ decay distribution  (where  $\ell$  is a charged lepton)  can be computed  in terms of the radiative branching fraction $ {\cal B}_{\rm rad} (M^\prime  \to M \gamma)$ and of a QED factor $F_{QED}$  which depends on  $q^2$, the squared  lepton pair invariant mass.
For extended hadrons  a transition form factor (TFF) must be included in the expression of the decay distribution:
\be
\frac{d {\cal B} (M^\prime  \to M  \ell^+ \ell^-)}{dq^2}={\cal B}_{\rm rad} (M^\prime  \to M \gamma)\,\frac{\lambda^{1/2}( m_{M^\prime}^2,m_M^2,q^2)}{\lambda^{1/2}( m_{M^\prime}^2,m_M^2,0)}\,  F_{QED}(q^2)  \,\, |f(q^2)|^2 \,\,\, ,\label{generaleq}
\ee
with $\lambda$  the triangular function. $f(q^2)$ is the transition form factor normalized to $f(q^2=0)=1$  \cite{Faessler:1999de,Luchinsky:2017pby}. It is sensitive to the structure of the particles involved in the decay and  must be determined in a nonperturbative QCD approach or  by comparing the  spectrum to measurement. 

For heavy quarkonia,   the symmetries that emerge in the large heavy quark (HQ) mass limit establish relationships among the decay modes of different states. These symmetries allow us to utilise  existing measurements to predict the rates of unobserved decays. Furthermore, the connections of the Dalitz rates to the radiative ones 
are useful to gaining additional information about   $\chi_{c1}(3872)$ meson through Dalitz processes. The relationship involves the measured radiative  branching fraction ${\cal B}_1={\cal B}(\chi_{c1}(3872) \to J/\psi \gamma)$, as well as ${\cal B}_2={\cal B}(\chi_{c1}(3872) \to\psi(2S) \gamma)$ and their ratio ${\cal B}_2/{\cal B}_1$.  The availability of both the electron and muon channels   doubles the set of accessible observables.

We discuss the application of the heavy quark symmetries to  heavy quarkonia to determine  radiative and Dalitz decay amplitudes  in Section~\ref{sec2}, after a derivation of
Eq.~\eqref{generaleq}.  Section~\ref{numerics}  presents an analysis of charmonia and bottomonia Dalitz modes  with predictions relevant for  ongoing and future experimental investigations.  Section~\ref{darkphoton} contains a study of  the potential contribution of a light new vector mediator in two specific charmonium Dalitz channels   and an assessment of the precision required to achieve  sensitivity to this effect.  We conclude with a summary.

\section{ Dalitz decays of heavy quarkonia}\label{sec2}

\subsection{Dilepton mass distribution in Dalitz decays}
To apply  Eq.~\eqref{generaleq}  to heavy quarkonia  exploiting the relation between   Dalitz and  radiative modes,  we derive  this expression  in a straightforward way \cite{Faessler:1999de}.
Consider the decay $M^\prime (p^\prime=m_{M^\prime}v^\prime) \to M(p=m_M v) \ell^-(k_1) \ell^+(k_2)$ where $M^\prime$ and $M$ are heavy quarkonium states of mass $m_{M^\prime}$ and $m_M$ and  four-velocity $v^\prime$ and $v$, respectively. The internal photon  conversion amplitude 
\bea
{\cal A}(M^\prime \to M \ell^- \ell^+)&=&(i e \,e_Q) \langle M(v) |  {\bar Q} \gamma^\mu Q |  M^\prime (v^\prime) \rangle \frac{- ig_{\mu \nu}}{q^2}(-i e)\,{\bar u}_\ell(k_1) \gamma^\nu  v_\ell (k_2)
\nn \\
&=& - i e^2 \,e_Q \, {\cal M}^\mu \frac{g_{\mu \nu}}{q^2} L^\nu \,\, 
\eea
involves the matrix element  ${\cal M}^\mu$  of the quark vector current 
 ${\cal M}^\mu=\langle M(v) |{\bar Q} \gamma^\mu Q |M^\prime (v^\prime) \rangle$, with $Q$  the heavy quark of charge $e e_Q$  ($e$  is the positron charge).  $q=p^\prime-p$ is  the virtual photon momentum and $L^\nu$  the leptonic current. 
The $q^2$ distribution for a  decaying particle  of spin $J$ reads
\be
\frac{d\Gamma}{dq^2}(M^\prime \to M \ell^- \ell^+)=\frac{1}{(2J+1)}\,\frac{  \alpha e_Q^2}{4 m_{M^\prime}^3} \,
\lambda^{1/2}( m_{M^\prime}^2,m_M^2,q^2) |F(q^2)|^2\,\frac{\alpha}{3\pi q^4}(2m_\ell^2+q^2) \sqrt{1-\frac{4m_\ell^2}{q^2}} ,
\ee
with  $\alpha$  the fine structure constant, $|F(q^2)|^2= \sum {\cal M}^\mu  {\cal M^*}_\mu$ (the sum is over the hadron spins), and $m_\ell$ the lepton mass.
This can be written as 
\be
\frac{d\Gamma}{dq^2}(M^\prime \to M \ell^- \ell^+)=\Gamma(M^\prime \to M \gamma)\, \frac{\lambda^{1/2}( m_{M^\prime}^2,m_M^2,q^2)}{\lambda^{1/2}(m_{M^\prime}^2,m_M^2,0)}\, 
 \Big| \frac{F(q^2)}{F(0)} \Big|^2 \,\frac{\alpha}{3\pi q^4}(2m_\ell^2+q^2) \sqrt{1-\frac{4m_\ell^2}{q^2}} \, , 
\ee
where
\be 
\Gamma(M^\prime \to M \gamma)=\frac{1}{(2J+1)}\,\frac{ \alpha e_Q^2}{4 m_{M^\prime}^3} \,
\lambda^{1/2}( m_{M^\prime}^2,m_M^2,0) |F(0)|^2 \,\,
\ee
is the radiative decay width.
 Eq.~\eqref{generaleq} is  obtained defining $f(q^2)=\displaystyle\frac{F(q^2)}{F(0)}$ and
\be
F_{QED}(q^2)=\frac{\alpha}{3\pi q^4}(2m_\ell^2+q^2) \sqrt{1-\frac{4m_\ell^2}{q^2}}\,\,.\label{fQED}
\ee
The relationship between  Dalitz modes and radiative modes is useful to extract  information on  Dalitz processes using existing results on radiative modes, and vice-versa.
For heavy quarkonia  the analysis is further enhanced  by exploiting the  symmetries arising in QCD in the heavy quark limit.\footnote{Dalitz decays of open charm  mesons are discussed in \cite{Colangelo:2023iia}.}

\subsection{Heavy quark symmetries and radiative decays of quarkonia}\label{Lag-rad}
In the heavy quark limit the radiative decays of quarkonia, important for the description of  the Dalitz modes as witnessed by Eq.~\eqref{generaleq}, can be described in an effective Lagrangian approach. The effective Lagrangian is built on the basis of the symmetries  holding in QCD in the infinite mass limit $m_Q \to \infty$ (the heavy quark HQ limit)  \cite{Neubert:1993mb}.

The QCD Lagrangian in the HQ limit is obtained  defining the field 
$h_v(x)=e^{im_Q v \cdot x}P_+Q(x)$, where $Q$ is the HQ field in QCD and $v$ is the heavy quark  four-velocity. The field 
 $h_v$ is obtained through the velocity projector $P_+=\displaystyle{1 + \spur v \over 2}$,
$h_v=P_+ Q$, and satisfies the condition ${\spur v}h_v=h_v$. The QCD Lagrangian for the heavy quark Q can be written in terms of $h_v$ as an expansion in $m_Q^{-1}$, with the leading term in the expansion, (the heavy quark effective theory  (HQET) Lagrangian)  
  \be 
  {\cal L}_{HQET}={\bar h}_v i \, v \cdot D h_v \,\, . \label{hqetlag}
  \ee 
  $D$ is the QCD covariant derivative.   For $N_f$ heavy flavours the Lagrangian \eqref{hqetlag} is invariant under $SU(2 N_f)$ spin/flavour rotations. The next-to-leading terms in the heavy quark expansion  can be sistematically included,  and they break the spin/flavour symmetry.
The  subleading operators are suppressed by powers of $k/m_Q$, where $k$ is  a residual momentum of ${\cal O}(\Lambda_{QCD})$  introduced to take into account that the HQ is off shell  writing   its momentum  as $p=m_Q v+k$.
 At ${\cal O}(1/m_Q)$ two operators arise:
\be
{\cal L}^{(1)}={1 \over 2 m_Q} {\bar h}_v (i \spur
D_\perp)^2 h_v +{1 \over 2 m_Q}{\bar h}_v {g_s \sigma_{\alpha \beta}
G^{\alpha \beta} \over 2 } h_v 
\label{lag1m} \; ,
\ee
the  HQ kinetic energy  operator due to the residual momentum $k$,  and the  chromomagnetic operator due to coupling of the HQ spin to the gluon field. Both the operators break the flavour symmetry, and the second one also breaks the heavy quark spin symmetry. Such properties have extensively been applied in the analysis of hadrons comprising a single heavy quark  \cite{Isgur:1991wq}.
However, for  systems  comprising a heavy quark-antiquark pair, the heavy quarkonia,   the flavour symmetry cannot be exploited  even at the leading order in the expansion \cite{Thacker:1990bm}. The reason is that for two heavy quarks with the same velocity, diagrams describing their gluon exchanges are affected by infrared divergences that  can  be regulated  going beyond the leading order in the HQ expansion,  including of  the kinetic energy operator. 
Nevertheless, the HQ spin symmetry  still survives. 

A consequence of the spin symmetry is that in the heavy quark limit
hadrons  differing  for the HQ spin orientation  are  degenerate,  and they can be treated in a unified way  collecting them  in spin multiplets. The degeneracy is broken starting from terms of ${\cal O}(1/m_Q)$, i.e. from the chromomagnetic operator. Heavy-light mesons  can be organized in spin  doublets,  the two states obtained in correspondence to the two orientations of the  spin of the heavy quark.  For $Q{\bar Q}$ systems both the spin of the heavy quark and of   the heavy antiquark can be rotated,  and the  spin multiplets  contain a number of states which depends on the  relative orbital angular momentum $L$ between the heavy quarks. We are interested in  $L=0$ ($S$-wave) states, that can be organised in a spin doublet,  and $L=1$ ($P$-wave) states organised in a spin four-plet. The two multiplets  can be represented as follows \cite{Jenkins:1992nb,Casalbuoni:1996pg}:
\begin{itemize}
 \item L=0 doublet: 
 \be
J={ 1+ \spur{v} \over 2} \left[H_1^\mu \gamma_\mu -H_0 \gamma_5 \right]{ 1- \spur{v} \over 2} \,\,\label{Swave} 
\ee
\item L=1 four-plet:
 \be  
 J^{\mu }={ 1+ \spur{v} \over 2}\Big\{H_2^{\mu \alpha } \gamma_\alpha  + {1 \over \sqrt{2}} \epsilon^{\mu \alpha \beta \gamma} v_\alpha \gamma_\beta H_{1
\gamma}+ {1 \over \sqrt{3}} (\gamma^{\mu} -v^{\mu}) H_0  + K_1^{\mu }\gamma_5 \Big\}{ 1-\spur{v} \over 2}\,\,. \label{Pwave} 
\ee
\end{itemize}
 $v^\mu$ is  the heavy quark four-velocity, with  the transversality condition $v_\mu J^{\mu}=0$.
$H_A$ and $K_A$ are the effective fields for the members of the multiplets with total spin $J=A$. 

In terms of  the spin multiplets, effective Lagrangians can be constructed to describe the heavy quarkonium phenomenology.  We use 
 the effective Lagrangian describing radiative decays of heavy quarkonia  in the soft-gluon-exchange  approximation \cite{Casalbuoni:1992yd}.
The electric dipole radiative transitions of members of  $P-$wave  multiplet to members of the
$S-$wave doublet, with the orbital angular momentum of the decaying and of the produced quarkonium  differing by $\Delta L=1$, are described by the
the effective Lagrangian 
\be 
{\cal L}_{nP \leftrightarrow mS}=\delta^{nPmS}_Q {\rm Tr} \left[{\bar J}(mS) J_\mu(nP) \right] v_\nu F^{\mu \nu} + \rm{h.c.}
\,.\label{lagPS} 
\ee 
where $n,m$ are radial quantum numbers, $\bar J=\gamma^0 J^\dagger \gamma^0$ and
 $F^{\mu \nu}$ is the electromagnetic field strength tensor.  
This Lagrangian  is invariant under parity $(P)$,  charge conjugation $(C)$, and time reversal $(T)$ transformations:
\bea
 J^{\mu_1 \dots \mu_L}&& \stackrel{P}{\rightarrow}\gamma^0 J_{\mu_1 \dots \mu_L} \gamma^0
\label{parity} \\
J^{\mu_1 \dots \mu_L}&& \stackrel{C}{\rightarrow}(-1)^{L+1} C [J^{\mu_1 \dots \mu_L}]^T C
\label{charge-conj} \\
 J^{\mu_1 \dots \mu_L}&& \stackrel{T}{\rightarrow}-T J_{\mu_1 \dots \mu_L} T^{-1}
\label{time-reversal} 
\eea
where $C=i \gamma^2 \gamma^0$ and $T=i \gamma^1 \gamma^3$.  Moreover, 
the Lagrangian \eqref{lagPS} is invariant under heavy quark spin transformations, that for a generic  $L$-wave multiplet $J^{\mu_1 \dots \mu_L}$ read
\be 
J^{\mu_1 \dots \mu_L}
\stackrel{SU(2)_{S_h}}{\rightarrow}S J^{\mu_1 \dots \mu_L}
S^{\prime \dagger} \,\,.\label{HQspin} 
\ee 
 $S$, $S^\prime$ belong to $SU(2)_{S_h}$,  the group of heavy
quark spin rotations,  and satisfy the relations
 $[S,{\spur v}]=[S^\prime,{\spur v}]=0$.
The spin symmetry implies that the single coupling constant $\delta^{nPmS}_Q$
describes all     transitions between $Q \bar Q$  states in a  $nP$ multiplet to the states in a  $mS$ doublet \cite{Casalbuoni:1992yd,DeFazio:2008xq,Colangelo:2025uhs}.   The  expressions of the corresponding  decay widths are collected in the appendix \ref{appA}.

The $\Delta L=0$ magnetic dipole transitions among  members of the $S$-wave doublets are instead governed by the effective Lagrangian
\be
 {\cal L}_{nS \leftrightarrow mS}=\delta^{nSmS}_Q {\rm Tr} \left[{\bar J}(mS) \sigma_{\mu \nu}J(nS) \right]  F^{\mu \nu} + \rm{h.c.} \,, \label{lagSS} 
\ee 
which is  invariant under $P,\,C,\,T$ transformations but violates the spin symmetry 
\cite{Cho:1992nt}.  The resulting expression of the decay  width  \cite{Colangelo:2025uhs} is also reported in the appendix \ref{appA}.

The effective  Lagrangians   \eqref{lagPS} and \eqref{lagSS} can be exploited  to analyze the radiative  heavy quarkonia transitions,  to verify the accuracy of the HQ limit, to determine the couplings from the measured radiative rates and to predict unmeasured decay widths \cite{Colangelo:2025uhs}.  They can also be used to get information on the structure of debated states. For example,  the comparison of the  computed radiative decay rates  $\chi_{c1}(3872) \to \psi(1S,2S) \, \gamma$ with  measurements supports the identification of $\chi_{c1}(3872)$
as $\chi_{c1}(2P)$   \cite{Colangelo:2025uhs}. The framework based on  spin symmetry can  also be applied to the heavy quarkonium Dalitz modes.
\footnote{The radiative decay modes of  $h_{c,b}(1P)$ and  $\chi_{c1}(1P)$ are analysed by lattice QCD in  \cite{Becirevic:2025ocx,Becirevic:2025idm}.}

\section{Heavy quarkonium Dalitz modes}\label{numerics}
The $L=0$   charmonium and bottomonium  spin doublets are   $(\eta_c(nS),\psi(nS))$ and $(\eta_b(nS),\Upsilon(nS))$, with $J^{PC}=(0^{-+},1^{--})$ and  radial quantum number $n$.
For $L=1$ the spin multiplets comprise four states $(\chi_{c(b)0},\,\chi_{c(b)1},\,\chi_{c(b)2},\,h_{c(b)})$ with $J^{PC}=(0^{++},\,1^{++},\,2^{++},1^{+-})$.
The available  data quoted in \cite{ParticleDataGroup:2024cfk} for the radiative  and  Dalitz branching ratios  are collected in Table~\ref{BRccbar}. 
 Only in few cases   the rates of both channels are known.  The Dalitz branching ratios are not known for most  modes, in particular,
  no measurement is  available for  bottomonium  even though some processes have been observed. In our analysis we systematically make use of Eq.~\eqref{generaleq} relating the Dalitz decay distribution and the radiative branching fraction.

\begin{table}[t]
\centering 
\begin{tabular}{c c c c } 
\hline \hline
\small $M^\prime \to M$ & \small ${\cal B}(M^\prime \to M \gamma)$ & \small ${\cal B}(M^\prime \to M \mu^+ \mu^-)$ & \small ${\cal B}(M^\prime \to M e^+ e^-)$\\
\hline 
\small $\chi_{c0}(1P) \to J/\psi$ & \small $(1.41 \pm 0.09) \times 10^{-2} $ & \small  $<1.9 \times 10^{-5}$ &  \small $(1.34 \pm 0.30) \times 10^{-4}$\\
\small  $\chi_{c1}(1P) \to   J/\psi$ & \small $(34.3 \pm 1.3)  \times 10^{-2}  $  & \small $(2.33 \pm 0.29) \times 10^{-4}$&  \small $(3.46 \pm 0.24) \times 10^{-3}$\\
\small $\chi_{c2}(1P) \to  J/\psi$ & \small $(19.5 \pm 0.7) \times 10^{-2} $ &\small $(2.07 \pm 0.34) \times 10^{-4}$ & \small $(2.20 \pm 0.15) \times 10^{-3}$\\
\small $h_c(1P) \to  \eta_c $ & \small $(60 \pm 4) \times 10^{-2} $  &   &   \\ \\
\small $\psi(2S) \to \chi_{c0}(1P)$  & \small $(9.75 \pm 0.22) \times 10^{-2} $ &  & \small  $(1.05 \pm 0.25) \times 10^{-3}$ \\
\small $\psi(2S) \to  \chi_{c1}(1P)$  & \small $(9.75 \pm 0.27) \times 10^{-2} $   & & \small $(8.5 \pm 0.7) \times 10^{-4}$\\
\small $\psi(2S) \to  \chi_{c2}(1P)$  & \small $(9.38 \pm 0.23) \times 10^{-2} $ & & \small $(6.8 \pm 0.8) \times 10^{-4}$ \\ \\
\small $ \chi_{c1}(3872) \to J/\psi $ & \small  $ (7.8 \pm 2.9) \times 10^{-3} $  & & \\
\hline 
\small $\chi_{b0}(1P) \to \Upsilon(1S)  $ & \small $(1.94 \pm 0.27) \times 10^{-2} $ \\ 
\small $\chi_{b1}(1P) \to \Upsilon(1S)  $ & \small $(35.2 \pm 2.0) \times 10^{-2} $ \\ 
\small $\chi_{b2}(1P) \to \Upsilon(1S)  $ & \small $(18.0 \pm 1.0) \times 10^{-2} $ \\ 
\small $h_b(1P) \to \eta_b(1S) $ & \small $(52^{+6}_{-5}) \times 10^{-2} $ \\ \\
\small $\chi_{b0}(2P) \to \Upsilon(1S) $ & \small $(3.8 \pm 1.7) \times 10^{-3}$  \\ 
\small $\chi_{b1}(2P)\to \Upsilon(1S) $ & \small  $(9.9 \pm 1.0 ) \times 10^{-2} $  \\
\small $\chi_{b2}(2P)\to \Upsilon(1S) $ & \small $(6.6 \pm 0.8 ) \times 10^{-2} $ \\
\small $h_b(2P)\to \eta_b(1S) $ & \small $(22 \pm 5) \times 10^{-2} $ \\ 
\hline \hline
\end{tabular} 
\caption{\baselineskip 10pt  \small  Measured branching fractions of  radiative and Dalitz decays of charmonium and bottomonium states \cite{ParticleDataGroup:2024cfk}.  }
\label{BRccbar} 
\end{table}

\subsection{Charmonium $1P \to 1S$  Dalitz decays}\label{sec1P1S}
The transitions  from the charmonium $1P$ four-plet to the $1S$ doublet,  for which more abundant data are available, can be used to  verify  the results based on the heavy quark limit and to predict  the  branching fractions of unmeasured  Dalitz modes. The missing information is about  the transition form factor $f(q^2)$ that should be computed by nonperturbative QCD methods or using models \cite{Wang:2019fjb}.  An alternative approach, that we follow here, is to  bound the TFF using data. Inspired by vector meson dominance arguments, we choose  a  simple pole parametrization 
\be
f(q^2)=\frac{1}{1-\displaystyle\frac{q^2}{a^2}}\,\, \label{fpole}
\ee
which involves the mass parameter $a$. We assume this form in the full kinematical range of the momentum transferred to the lepton pair, 
$4m_\ell^2 \leq q^2 \leq (m_{M^\prime}-m_M)^2$.
Integrating  the distribution $\displaystyle\frac{d{\cal B}}{dq^2}$  in \eqref{generaleq}  one obtains the Dalitz branching fraction  depending on  ${\cal B}_{\rm rad}$ and on the mass parameter $a$.
The linear dependence on the radiative ${\cal B}_{\rm rad}$  gives the excursion for the Dalitz branching ratio:
${\cal B}_{\rm min}(M^\prime \to M \, \ell^+ \ell^-)={\cal B}(a,\,{\cal B}_{\rm rad} - \Delta {\cal B}_{\rm rad})$,
${\cal B}_{\rm max }(M^\prime \to M \, \ell^+ \ell^-)={\cal B}(a,\,{\cal B}_{\rm rad} + \Delta {\cal B}_{\rm rad})$,
 with $\Delta {\cal B}_{\rm rad}$ the error on ${\cal B}_{\rm rad}$.
 
The radiative widths  computed using the effective Lagrangian \eqref{lagSS} and collected in  appendix \ref{appA},   when compared to measurement,  confirm that  the single coupling  $\delta_{1P1S}$   governs the $1P \to 1S$ transitions   for all members of the initial  four-plet decaying to states in the final doublet \cite{Colangelo:2025uhs}.
In the heavy quark  limit the matrix elements of the  current ${\bar Q} \Gamma Q$ (with $\Gamma$  a Dirac matrix)  between two quarkonium states for $q^2$ close to $q^2_{\rm max}$   can be related   \cite{Jenkins:1992nb}. Such relations can be extended  to the full kinematic range for the narrow phase space of the  Dalitz decays to muons.  Moreover, in the HQ limit the mass splitting among the states belonging the various multiplets vanishes.
The  consequence is that the mass parameter $a$ in \eqref{fpole} is the same for all $1P \to 1S$ transitions.

The value of $a$ can be obtained from data. For the transitions $\chi_{c1}(1P) \to J/\psi $ and $\chi_{c2}(1P) \to J/\psi$, both the radiative and the Dalitz branching fractions  are measured, and $a$ is constrained requiring that the range  $[{\cal B}_{\rm min}(M^\prime \to M \, \ell^+ \ell^-),\,{\cal B}_{\rm max}(M^\prime \to M \, \ell^+ \ell^-)]$ overlaps with the experimental range in Table~\ref{BRccbar}. 
The obtained intervals are  further restricted minimizing   $\chi^2=\displaystyle{({\cal B}(a)-{\cal B}_{\rm exp})^2}/{\sigma_{\exp}^2}$, with ${\cal B}(a)$  the prediction for ${\cal B}(M^\prime \to M \,\mu^+ \mu^-)$ obtained for  the central value of the radiative ${\cal B}_{\rm rad}$,  ${\cal B}_{\rm exp}$ and $\sigma_{\exp}$ the Dalitz branching fraction and error in Table \ref{BRccbar}.  We use the meson masses  quoted in \cite{ParticleDataGroup:2024cfk}. From the mode  $\chi_{c1}(1P) \to J/\psi \, \mu^+ \mu^-$ we obtain
  $a_{c1}=0.83^{+0.17}_{-0.11}$ GeV,   from $\chi_{c2}(1P) \to J/\psi \, \mu^+ \mu^-$ we have $a_{c2}=0.71^{+0.13}_{-0.07}$ GeV. 
As expected,   the pole mass parameters  in the two modes are close to each other.  In correspondence to  $a_{c1}$ and  $a_{c2}$ the computed branching fractions  of the muon and electron modes are
\bea
{\cal B}(\chi_{c1}(1P) \to J/\psi \, \mu^+ \mu^- (e^+ e^-) )&=&(2.34 \pm 0.10)\times 10^{-4} \hskip 0.3cm ((3.06 \pm 0.12)\times 10^{-3} )\\
{\cal B}(\chi_{c2}(1P) \to J/\psi \,\mu^+ \mu^- (e^+ e^-) )&=&(2.06 \pm 0.10)\times 10^{-4}  \hskip 0.3cm ((1.83 \pm 0.08)\times 10^{-3} ) \,\, .
\eea
Due to the larger  phase space for  the dielectron channel,  the extension of the HQ limit  to the whole kinematical range is expected to produce a less satisfactory agreement with  measurements in case of  electrons. Using together  the measured ${\cal B}(\chi_{c1}(1P) \to J/\psi \, \mu^+ \mu^-)$ and  ${\cal B}(\chi_{c2}(1P) \to J/\psi \, \mu^+ \mu^-)$   we obtain   \be
  a_c=0.77 \pm 0.10 \,\, {\rm GeV}\, ,  \label{ac}
  \ee
the value  we use for all $1P \to 1S$ transitions.

The computed branching fractions  are collected in Table \ref{BRccbarth};  some of them are predictions for unobserved $1P \to 1S$ charmonium modes.
The uncertainties quoted in Table \ref{BRccbarth} are mainly due to the errors for the branching rations of the radiative decays, which cancel out in the ratios between the muonic and the electronic  decay rates. For this reason, in Table \ref{BRccbarth} we do not quote the error for  $\dd \frac {{\cal B}(M^\prime \to M \mu^+ \mu^-)}{{\cal B}(M^\prime \to M e^+ e^-)}$. The BESIII measurements  
$\dd \frac {{\cal B}(\chi_{c1} \to J/\psi \mu^+ \mu^-)}{{\cal B}(\chi_{c1} \to J/\psi e^+ e^-)}=(6.73 \pm 0.51\pm 0.50)\times 10^{-2}$, 
$\dd\frac {{\cal B}(\chi_{c2} \to J/\psi \mu^+ \mu^-)}{{\cal B}(\chi_{c2} \to J/\psi e^+ e^-)}=(9.40 \pm 0.79\pm 1.15)\times 10^{-2}$, 
and   the upper bound
$\dd\frac {{\cal B}(\chi_{c0} \to J/\psi \mu^+ \mu^-)}{{\cal B}(\chi_{c0} \to J/\psi e^+ e^-)} < 14\times 10^{-2}$ (at 90 \% C.L.) \cite{BESIII:2019yeu},  in which the systematic uncertainties largely cancel out,    agree with the results in  Table \ref{BRccbarth}. 
 \begin{table}[t!]
\centering 
\begin{tabular}{ c c c c } 
\hline \hline
\small $M^\prime \to M$ &  \small ${\cal B}(M^\prime \to M \mu^+ \mu^-)$ & \small  $\dd {\cal B}(M^\prime \to M e^+ e^-)$& \small $\frac {{\cal B}(M^\prime \to M \mu^+ \mu^-)}{{\cal B}(M^\prime \to M e^+ e^-)}$\\
\hline 
\small $\chi_{c0}(1P) \to J/\psi$ & \small $(4.02\pm 0.26)\times 10^{-6}$ & \small  $(1.2\pm 0.1)\times 10^{-4}$& \small $3.4 \times 10^{-2}$ \\
\small $\chi_{c1}(1P) \to J/\psi$ & \small $(2.47 \pm 0.09) \times 10^{-4}$ & \small  $(3.08 \pm 0.12)\times 10^{-3}$& \small $8.0 \times 10^{-2}$\\
\small $\chi_{c2}(1P) \to J/\psi$ & \small $(1.85 \pm 0.08)\times 10^{-4}$ & \small $(1.80 \pm 0.08)\times 10^{-3}$& \small $10.3 \times 10^{-2}$\\
\small $h_{c}(1P) \to \eta_c$ & \small $(8.7 \pm 0.6)\times 10^{-4}$ & \small $(5.9 \pm 0.4)\times 10^{-3}$& \small $14.7 \times 10^{-2}$\\
\hline \hline
\end{tabular} 
\caption{\baselineskip 10pt  \small  Computed branching fractions of $1P \to 1S$ charmonium  Dalitz  modes using the TFF in \eqref{fpole} and the  pole mass parameter in \eqref{ac}.}
\label{BRccbarth} 
\end{table}

For    ${\cal B}(\chi_{c0}(1P) \to J/\psi \, e^+ e^-)$ the obtained value favourably compares with the average  in Table~\ref{BRccbar}.
The  prediction for ${\cal B}(\chi_{c0}(1P) \to J/\psi \,\mu^+ \mu^-)$ is  also compatible with the upper bound reported in the same Table.
For the   Dalitz decays of $h_c(1P)$ no result is quoted in \cite{ParticleDataGroup:2024cfk}. An analysis by the BESIII Collaboration reports the measurement of the ratio 
${\cal R}=\displaystyle\frac{{\cal B}(h_c(1P) \to \eta_c(1S) \, e^+ e^-)}{{\cal B}(h_c(1P) \to \eta_c(1S)  \, \gamma)}$, with result depending on the  $h_c$  production mechanism \cite{BESIII:2024kkf}. Two possibilities are considered:
 $h_c$  producted via  $\psi (3686) \to e^+ e^- h_c$,
which  gives ${\cal R}=(0.46 \pm 0.12 \pm 0.05) \times 10^{-2}$, and  $h_c$ produced through $e^+ e^- \to \pi^+ \pi^- h_c$, 
which gives  ${\cal R}=(0.89 \pm 0.19 \pm 0.09) \times 10^{-2}$.
Combining the two results, the  average is quoted:  ${\cal R}=(0.59 \pm 0.10 \pm 0.04) \times 10^{-2}$,
 where the first error is statistical, the second systematic.
Using the radiative branching fraction in Table~\ref{BRccbar}, this  corresponds to
${\cal B}(h_c(1P) \to \eta_c(1S) \, e^+ e^-)=(3.5 \pm 0.9)\times 10^{-3}$\,\,.
The prediction  in Table~\ref{BRccbarth} is marginally consistent with  this experimental result, 
which however suffers from the  method to obtain it. 

\begin{figure}[t!]
\begin{center}
\includegraphics[width = 0.44\textwidth]{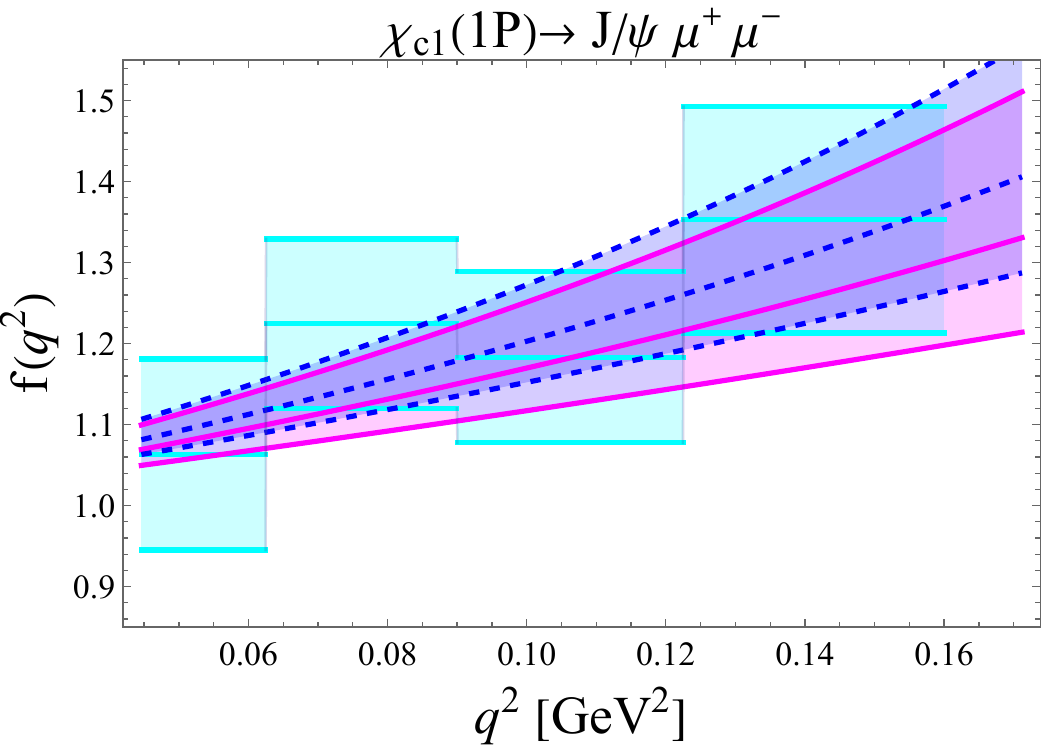} 	\hskip 0.5cm
\includegraphics[width = 0.44\textwidth]{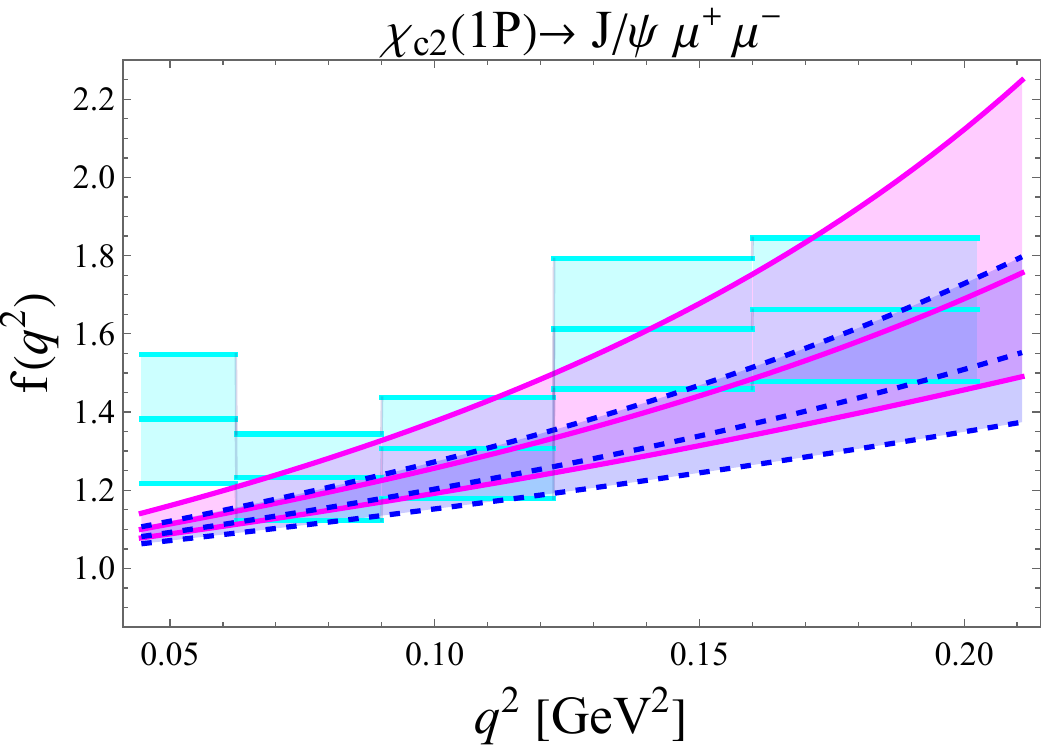}
\caption{\baselineskip 10pt  \small  Transition form factor $f(q^2)$ for  $\chi_{c1}(1P) \to J/\psi \mu^+ \mu^-$ (left panel) and   $\chi_{c2}(1P) \to J/\psi \mu^+ \mu^-$ (right panel). The binned cyan regions are the BESIII results  \cite{BESIII:2019yeu}. The magenta regions are obtained using Eq.~\eqref{fpole} with  the values  $a_{c1}$ and $a_{c2}$ determined independently for the two modes, the blue regions   are obtained for the common value $a_c$ in \eqref{ac}.}\label{figFF}
\end{center}
\end{figure}

For  the modes $\chi_{c1}(1P) \to J/\psi \mu^+ \mu^-$ and $\chi_{c2}(1P) \to J/\psi \mu^+ \mu^-$  the BESIII Collaboration  has    measured the branching fraction respectively  in four and five bins of $q^2$,  determining in each bin the transition form factor  \cite{BESIII:2019yeu}. In Fig.~\ref{figFF} we compare the result obtained using  $a_{c1}$,  $a_{c2}$ (magenta regions) and $a_c$ (dark blue regions) to the  BESIII  measurement (binned cyan regions).   
The effect of the transition form factor also emerges considering the ratios
$\dd {\cal R}^{ee}_{J,\gamma}=\frac {{\cal B}(\chi_{cJ} \to J/\psi e^+ e^-)}{{\cal B}(\chi_{cJ} \to J/\psi \gamma)}$. With the expression  \eqref{fpole} we obtain 
 ${\cal R}^{ee}_{0,\gamma}=8.4 \times 10^{-3}$, ${\cal R}^{ee}_{1,\gamma}=9.0 \times 10^{-3}$, ${\cal R}^{ee}_{2,\gamma}=9.2 \times 10^{-3}$, to be
  compared to the  measurements
${\cal R}^{ee}_{0,\gamma}|_{exp}=(9.5 \pm 1.9\pm0.7) \times 10^{-3}$,
${\cal R}^{ee}_{1,\gamma}|_{exp}=(10.1 \pm 0.3 \pm 0.5) \times 10^{-3}$,
${\cal R}^{ee}_{2,\gamma}|_{exp}=(11.3 \pm 0.4 \pm 0.5) \times 10^{-3}$  \cite{BESIII:2017ung}.  For  constant TFF, $f(q^2)=1$, the computed ratios are systematically reduced:
${\cal R}^{ee}_{0,\gamma}=8.2 \times 10^{-3}$, ${\cal R}^{ee}_{1,\gamma}=8.6 \times 10^{-3}$, ${\cal R}^{ee}_{2,\gamma}=8.8 \times 10^{-3}$.

\begin{figure}[b!]
\begin{center}
\includegraphics[width = 0.44\textwidth]{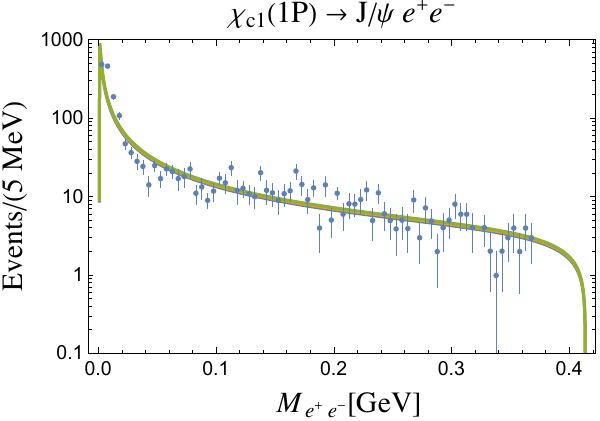}  \hskip 0.5cm 
\includegraphics[width = 0.44\textwidth]{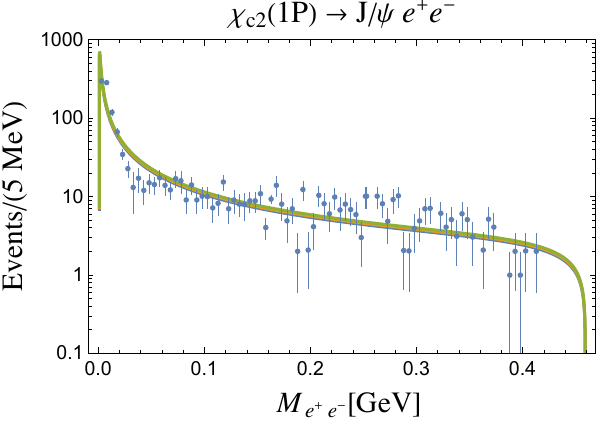} 
\caption{\baselineskip 10pt  \small  $M_{e^+ e^-}$ distributions  for  $\chi_{cJ}(1P) \to J/\psi \, e^+ e^-$   ($J=1,2$) in the full kinematical range,  computed using the TFF in \eqref{fpole} and the  pole mass parameter in \eqref{ac} (continuous line), compared to the   BESIII  data  \cite{BESIII:2017ung}. }\label{spectraeefull}
\end{center}
\end{figure}

\begin{figure}[t!]
\begin{center}
\includegraphics[width = 0.44\textwidth]{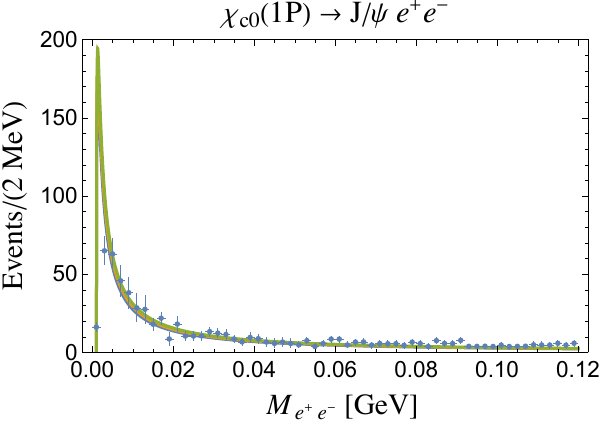}   \\ \vskip 0.5cm
\includegraphics[width = 0.44\textwidth]{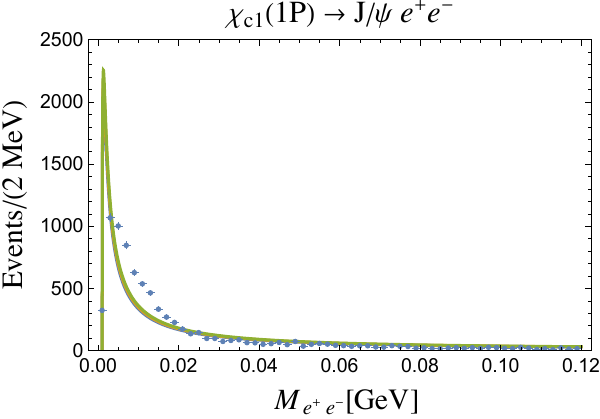}  \hskip 0.5cm 
\includegraphics[width = 0.44\textwidth]{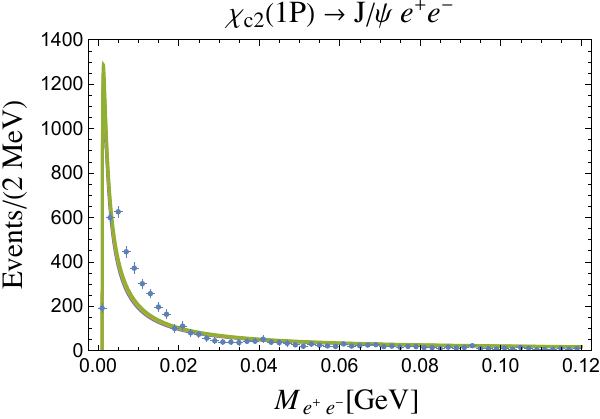} 
\caption{\baselineskip 10pt  \small  $M_{e^+ e^-}$ distributions  for  $\chi_{cJ}(1P) \to J/\psi \, e^+ e^-$   ($J=0,1,2$)  in the low range of dilepton mass, computed using the TFF in \eqref{fpole} and the  pole mass parameter in \eqref{ac} (continuous line).  The dots are the  BESIII  measurement  \cite{BESIII:2025otp}. }\label{spectraee}
\end{center}
\end{figure}

The $M_{e^+ e^-}$ distributions of  $\chi_{cJ}(1P) \to \psi(1S) \, e^+ e^-$ have been measured in the full kinematic range  for $J=1,2$ \cite{BESIII:2017ung}, and recently with higher precision  in the low range up to $0.12$ GeV for $J=0,1,2$ \cite{BESIII:2025otp}.  Figures~\ref{spectraeefull} and \ref{spectraee} show the  comparison of the  distributions obtained using Eqs.~\eqref{generaleq}, \eqref{fpole} and  \eqref{ac} (setting a suitable normalization in the various cases) with the measured spectra. An overall agreement is found. The enhancement near the real photon production is clearly observed.
 In the case of $\chi_{c1}(1P)$  and $\chi_{c2}(1P)$ there is an apparent excess of events in the  dielectron mass  distribution for low $M_{e^+ e^-}$, see Fig.~\ref{spectraee}. 
 This is a warning that we shall discuss in the last Section.
 
\subsection{Charmonium $2S \to 1P$  Dalitz  decays}
\begin{figure}[h!]
\begin{center}
\includegraphics[width = 0.44\textwidth]{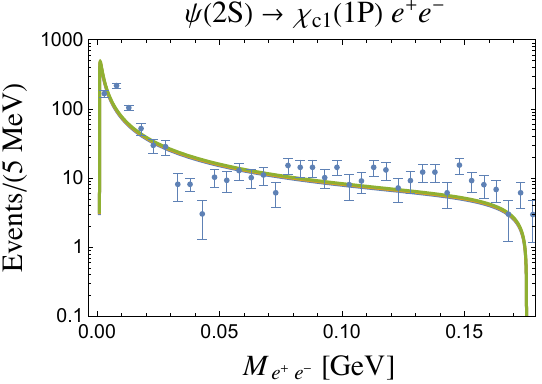}  \hskip 0.5cm 
\includegraphics[width = 0.44\textwidth]{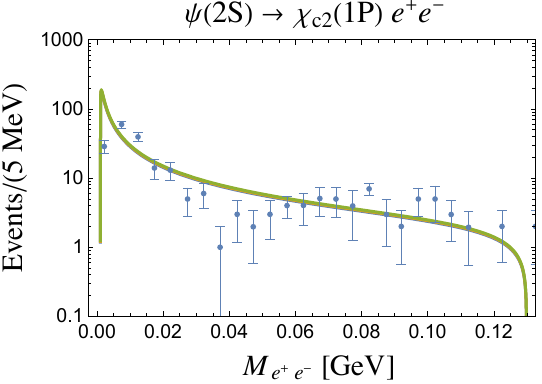} 
\caption{\baselineskip 10pt  \small  $M_{e^+ e^-}$ distributions  of  $\psi(2S) \to \chi_{cJ}(1P) \, e^+ e^-$   (for $J=1,2$) in the full kinematical range, computed using the TFF in \eqref{fpole} with the  parameter in \eqref{ac2S} (continuous line), compared to the 
  BESIII  data  \cite{BESIII:2017ung}. }\label{spectrapsi2Schicee}
\end{center}
\end{figure}

Among the transitions from the $2S$ doublet to the $1P$ four-plet,
measurements are available for the widths of the decays  $\psi(2S) \to \chi_{cJ} \gamma$ and  of the  Dalitz  modes $\psi(2S) \to \chi_{cJ} \, e^+ e^-$ $(J=0,1,2)$ in  Table~\ref{BRccbar}.  No data are available for  Dalitz decays to muons.  
The measured   ${\cal B}(\psi(2S) \to \chi_{cJ} \, e^+ e^-)$ can be used  to determine the  TFF mass parameter:
\be
a_{c,2S}=0.286^{+0.009}_{-0.001}  \,\, {\rm GeV} \,\, .\label{ac2S}
\ee
The comparison of the   $M_{e^+ e^-}$ distribution obtained with this  TFF, setting    a suitable normalization,   with the spectra measured by BESIII \cite{BESIII:2017ung},  shown in Fig.~\ref{spectrapsi2Schicee},  supports  the statement that the same transition form factor  enters in both  the modes $\psi(2S) \to \chi_{c1,2}(1P) \, e^+ e^-$.  The small  value obtained for $a_{c,2S}$ is responsible of the agreement mainly  for  large  dielectron invariant mass. 

With the parameter  \eqref{ac2S} we compute the branching fractions  in Table~\ref{psi2Sll}.
 \begin{table}[h!]
\centering 
\begin{tabular}{ c c c } 
\hline \hline
\text{\small $M^\prime \to M$} &  \small ${\cal B}(M^\prime \to M \mu^+ \mu^-)$ & \small ${\cal B}(M^\prime \to M e^+ e^-)$\\
\hline 
\small $\psi(2S) \to \chi_{c0}(1P)$ & \small $(1.51 \pm 0.04 )\times 10^{-4}$ & \small $(11.1 \pm 0.3) \times 10^{-4}$\\
\small $\psi(2S) \to \chi_{c1}(1P)$ & \small $(2.46 \pm 0.07) \times 10^{-5}$ & \small $(7.6 \pm 0.2) \times 10^{-4}$\\
\small $\psi(2S) \to \chi_{c2}(1P)$ & \small $(1.53 \pm 0.04 )\times 10^{-4}$ & \small $(6.6 \pm 0.2) \times 10^{-4}$\\
\small $\eta_c(2S) \to h_c(1P)$ & \small $(5.7 \pm 0.9) \times 10^{-6}$ & \small $(1.35\pm 0.2) \times 10^{-5}$\\ \\
\small $\chi_{c1}(2P) \to J/\psi$ & \small $ (4.2 \pm 1.7) \times 10^{-5}$ & \small $(1.3 \pm 0.5) \times 10^{-4} $ \\
\small $\chi_{c2}(2P) \to J/\psi$ & \small $ (1.6 \pm 0.7) \times 10^{-5}$ & \small $(2.9 \pm 0.9) \times 10^{-5} $ \\
\hline \hline
\end{tabular} 
\caption{\baselineskip 10pt  \small  Branching fractions of charmonium  $2S \to 1P$ and $2P\to 1S$  Dalitz decays computed using spin symmetry relations. For   $\chi_{c1}(2P)$ see the discussion in the text. For $\chi_{c0}(2P)$ and $h_c(2P)$, the uncertainties of the input data (when available) are too  large to obtain predictions.}
\label{psi2Sll} 
\end{table}
For  $\eta_c(2S)$  no measurement is available.  Using the expressions in appendix \ref{appA},  the HQ spin symmetry relations allowed us to predict
${\cal B}(\eta_c(2S) \to h_c(1P) \, \gamma)=(0.20\pm 0.03)\times 10^{-2}$  \cite{Colangelo:2025uhs}. 
With this result and the mass parameter  in \eqref{ac2S} we  obtain the predictions for  ${\cal B}(\eta_c(2S) \to h_c(1P) \mu^+ \mu^-)$ and ${\cal B}(\eta_c(2S) \to h_c(1P) e^+ e^-)$  in  Table~\ref{psi2Sll}.

\subsection{Charmonium $2P \to 1S$ Dalitz  decays}
The $2P \to 1S$ transitions are of interest since they  involve the debated state $\chi_{c1}(3872)$.  If  $\chi_{c1}(3872)$ is identified with $\chi_{c1}(2P)$,
piece of information comes from the measured   ${\cal B}(\chi_{c1}(3872) \to J/\psi \gamma)$  in  Table~\ref{BRccbar}. 
 Exploting the HQ relations, using meson masses and widths in  \cite{ParticleDataGroup:2024cfk}  and  the expressions in appendix \ref{appA},   this measurement implies\footnote{This value updates the result in  \cite{Colangelo:2025uhs} due to a slight change of the datum for ${\cal B}(\chi_{c1}(3872) \to J/\psi \gamma)$ used as  input.}
\be
{\cal B}(\chi_{c2}(2P) \to J/\psi \gamma)=(4.0 \pm 1.8) \times 10^{-4} \label{chic22P}\,\,.
\ee
In the HQ limit the TFF mass parameter $a$   is expected to be the same for 
$\chi_{c1}(2P) \to J/\psi \ell^+ \ell^-$ and $\chi_{c2}(2P) \to J/\psi \ell^+ \ell^-$. 
Imposing a conservative bound 
\be
0.3 \le\displaystyle\frac{{\cal B}(\chi_{c2}(2P) \to J/\psi \mu^+ \mu^-)}{{\cal B}(\chi_{c1}(2P) \to J/\psi \mu^+ \mu^-)}\le 1 \,\, , \label{ratio2P}
\ee
we obtain 
\be
a_{c,2P}=0.826 \,\, {\rm GeV} .
\ee
 This allows us to  compute  the branching fractions of the Dalitz modes  $\chi_{c1,2}(2P) \to J/\psi$ reported in Table~\ref{psi2Sll}.
 For the  $\chi_{c0}(2P)$ and $h_c(2P)$ Dalitz modes, the input data (when available) are too uncertain to derive predictions. 
 In the case of $\chi_{c1}(2P)$ decay, a further contribution  to the $\chi_{c1}(2P) \to J/\psi \ell^+ \ell^-$  amplitude comes from $\chi_{c1}(2P) \to J/\psi (\rho^0, \omega) \to J/\psi \ell^+ \ell^-$. It involves the isospin-violating $\chi_{c1}(2P) \to J/\psi \rho^0$ mode observed for $\chi_{c1}(3872)$, the branching fraction of which is quoted in \cite{ParticleDataGroup:2024cfk} and is connected  to the tale of the $\rho^0$ line-shape 
 (the intermediate  virtual $\omega$ is too narrow for providing a relevant effect). In order to estimate such a contribution, we compute the $\chi_{c1}(2P)  \to J/\psi \ell^+ \ell^-$ branching fraction taking into account only the $\rho^0$-exchange amplitude. We use  a constant transition form factor and a $\rho^0$ mass $0.5$ MeV away from the nominal value. We obtain results about a factor of 2 smaller than those quoted in  Table~\ref{psi2Sll}, both for the electron and muon mode. This implies that this contribution cannot be neglected, mainly because the relative phase  with respect to the internal photon conversion amplitude is not known.
 A possiblity to account for this contribution is to double the errors quoted in  Table~\ref{psi2Sll}. The study of the $q^2$ distribution would allow to isolate the  vector meson-exchange  contribution, since an  enhancement is foreseen in the end-point $q^2$ region. 

\subsection{ Bottomonium $1P \to 1S$ and $2P \to 1S$ Dalitz  modes}

\begin{figure}[t!]
\begin{center}
\includegraphics[width = 0.44\textwidth]{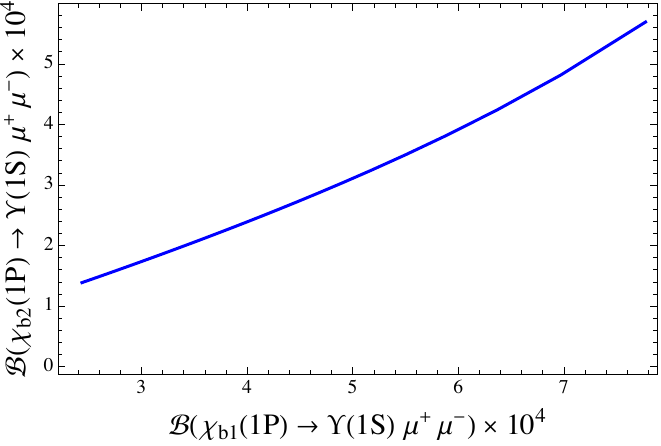} \hskip 0.5cm
\includegraphics[width = 0.44\textwidth]{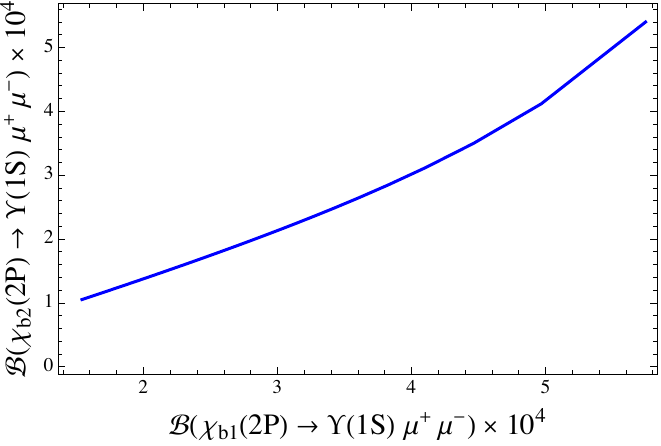}
\caption{\baselineskip 10pt  \small  Correlation  between  ${\cal B}(\chi_{b2}(1P)  \to \Upsilon(1S) \,\mu^+ \mu^-)$ and ${\cal B}(\chi_{b1}(1P)  \to \Upsilon(1S) \,\mu^+ \mu^-)$ (left panel) and between  ${\cal B}(\chi_{b2}(2P)  \to \Upsilon(1S) \,\mu^+ \mu^-)$ and ${\cal B}(\chi_{b1}(2P)  \to \Upsilon(1S) \,\mu^+ \mu^-)$ (right panel), obtained varying the TFF  parameters $a_{b,1P}$ and $a_{b,2P}$   in the ranges in Eq.~\eqref{ab}.}\label{figRatiosb}
\end{center}
\end{figure}

For bottomonium  there are   no measurements of  $\chi_{bJ}$ and $h_b$ Dalitz decay widths.   Piece of information 
has been  provided by the LHCb  Collaboration with the analysis of  $\chi_{b1,2}(1P)  \to \Upsilon(1S) \,\mu^+ \mu^-$  and of the analogous modes for $2P$ excitations $\chi_{b1,2}(2P)  \to \Upsilon(1S) \,\mu^+ \mu^-$  \cite{LHCb:2024rjv}.   The analysis has improved the  mass determination of 
$\chi_{b1,2}(1P,2P)$  with respect to the PDG values:
$m_{\chi_{c1}(1P)}=9892.50\pm0.26\pm0.10\pm0.10$ MeV,
$m_{\chi_{c2}(1P)}=9911.92\pm0.29\pm0.11\pm0.10$ MeV,
$m_{\chi_{c1}(2P)}=10253.97\pm0.75\pm0.22\pm0.09$ MeV,
$m_{\chi_{c2}(2P)}=10269.67\pm0.67\pm0.22\pm0.09$ MeV
(the first error  is statistical, the second error systematic,  the third one from the knowledge of the $\Upsilon(1S)$ mass) \cite{LHCb:2024rjv}.  Moreover,
 yields has been measured from a fit of the $\Upsilon(1S) \mu^+ \mu^-$ mass distribution:
\bea
N(\chi_{b1}(1P)  \to \Upsilon(1S) \,\mu^+ \mu^-) &=& 53.6 \pm 7.7 \, ,\quad \,\,\, 
N(\chi_{b2}(1P)  \to \Upsilon(1S) \,\mu^+ \mu^-) = 47.9 \pm 7.4 \, , \nn \\
N(\chi_{b1}(2P)  \to \Upsilon(1S)\, \mu^+ \mu^-) &=& 51.1 \pm 10.4  \, , \quad
N(\chi_{b2}(2P)  \to \Upsilon(1S) \,\mu^+ \mu^-) = 59.3 \pm 10.4  \, . \qquad 
\eea
 For  each process, the yield $N$  is proportional to the decay branching fraction ${\cal B}$ through the relation
\be
N=L \times \sigma \times A \times \epsilon \times {\cal B} \,\,\, ,  \label{yield}
\ee
where $L$ is the integrated luminosity, $A$ the acceptance, $\sigma$ the production cross section and $\epsilon$  the reconstruction efficiency.
We consider  ratios of branching fractions for decaying particles in the same spin multiplet:
\bea
R_{b,1P}&=&\frac{{\cal B}(\chi_{b2}(1P)  \to \Upsilon(1S) \,\mu^+ \mu^-)}{{\cal B}(\chi_{b1}(1P)  \to \Upsilon(1S) \,\mu^+ \mu^-)} \label{Rb1}
\\
R_{b,2P}&=&\frac{{\cal B}(\chi_{b2}(2P)  \to \Upsilon(1S) \,\mu^+ \mu^-)}{{\cal B}(\chi_{b1}(2P)  \to \Upsilon(1S)\, \mu^+ \mu^-)} \label{Rb2}
\eea
and,  invoking spin symmetry, we  assume that   for the two modes in each ratio \eqref{Rb1} and \eqref{Rb2} the TFF is the same.
 The quantities in Eq.~\eqref{yield} depending only on the detector and not on the decaying particle  cancel out  in the  ratios. As for the production cross section,  a simple argument   assumes  that it is proportional to the number of the spin states of the decaying particle.\footnote{Deviations are expected,  in particular  for production in the low $p_T$ region. We thank I.~Belyaev for pointing this out to us.}  This gives  $\sigma(\chi_{b2})/\sigma(\chi_{b1})\simeq 5/3$. As a result, we have
\bea
R_{b,1P}&\simeq &\frac{3}{5}\frac{N(\chi_{b2}(1P)  \to \Upsilon(1S) \,\mu^+ \mu^-) }{N(\chi_{b1}(1P)  \to \Upsilon(1S) \,\mu^+ \mu^-)} =0.54\pm0.11 \label{assumptionRb1} \\
R_{b,2P} &\simeq & \frac{3}{5}\frac{N(\chi_{b2}(2P)  \to \Upsilon(1S) \,\mu^+ \mu^-) }{N(\chi_{b1}(2P)  \to \Upsilon(1S) \,\mu^+ \mu^-)} = 0.70 \pm 0.19 \label{assumptionRb2} \,\,.
\eea
Requiring that the  ratios   \eqref{assumptionRb1} and \eqref{assumptionRb2} are recovered within $2\sigma$  constrains each TFF mass parameter $a$:
\be
a_{b,1P} \in [0.48, 1]  \,\,  {\rm GeV} \,\, ,  \hskip 1 cm a_{b,2P}\in [0.82, 2] \,\, {\rm GeV} \,\,.
\label{ab}
\ee 
Varying the mass parameter  $a$ in  \eqref{ab} produces a correlation between the two modes in each  ratio, as shown in Fig.~\ref{figRatiosb}.
Although the bounds \eqref{assumptionRb1} and \eqref{assumptionRb2} can be fulfilled for values of  $a_{b,1P}$ and $a_{b,2P}$ above the ranges  \eqref{ab},  there are no visible differences in the correlation plots since such large values  only populate the region of  smallest  branching fractions. 

Using  the radiative rates in Table~\ref{BRccbar} and setting  $a_{b,1P}=0.74$  GeV and $a_{b,2P}=1.41$ GeV, we compute the  branching ratios of all  $1P \to 1S$ and $2P \to 1S$ Dalitz bottomonium decays   in Table~\ref{BRbbbarll}. 
 The branching fractions of the Dalitz decays of  $\chi_{b1,2}(2P)$ due  to  $\omega$-exchanges, obtained using the measured ${\cal B}(\chi_{b1,2}(2P) \to Y(1S) \, \omega)$ \cite{ParticleDataGroup:2024cfk},  turn out to be  two orders of magnitude smaller than the ones quoted in  Table~\ref{BRbbbarll}. The $\omega$-exchange contribution  can be efficiently subtracted in the $q^2$ distributions. 
 The results  in Table~\ref{BRbbbarll}  are within the reach of  the present experimental facilities. 

\begin{table}[h!]
\centering 
\begin{tabular}{c  c  c} 
\hline \hline
\small $M^\prime \to M$& \small ${\cal B}(M^\prime \to M \mu^+ \mu^-)$  & \small ${\cal B}(M^\prime \to  M e^+ e^-)$  \\ 
\hline 
\small $\chi_{b0}(1P) \to \Upsilon(1S)$& \small $(1.3 \pm 0.2) \times 10^{-5}$ & \small $(1.7 \pm 0.25) \times 10^{-4}  $ \\ 
\small $\chi_{b1}(1P) \to \Upsilon(1S)$& \small $(3.0 \pm 0.2) \times 10^{-4} $ & \small $(3.2 \pm 0.2) \times 10^{-3}$ \\ 
\small $\chi_{b2}(1P) \to \Upsilon(1S)$& \small $(1.7 \pm 0.1) \times 10^{-4} $ & \small $(1.7 \pm 0.1) \times 10^{-3} $ \\ 
\small $h_{b}(1P) \to \eta_b(1S)$& \small $(6.5 \pm 0.8) \times 10^{-4} $ & \small $(5.6 \pm 0.6) \times 10^{-3} $ \\ \\
\small $\chi_{b0}(2P) \to \Upsilon(1S)$& \small $(6.4 \pm 2.9) \times 10^{-6}$ & \small $(3.8 \pm 1.7) \times 10^{-5}  $ \\ 
\small $\chi_{b1}(2P) \to \Upsilon(1S)$& \small $(1.75 \pm 0.2) \times 10^{-4} $ & \small $(1.0 \pm 0.1) \times 10^{-3}$ \\ 
\small $\chi_{b2}(2P) \to \Upsilon(1S)$& \small $(1.2 \pm 0.15) \times 10^{-4} $ & \small $(6.7 \pm 0.8) \times 10^{-4} $ \\ 
\small $h_{b}(2P) \to \eta_b(1S)$& \small $(4.35 \pm 1.0) \times 10^{-4} $ & \small $(2.25 \pm 0.5) \times 10^{-3} $ \\ 
\hline \hline
\end{tabular} 
\caption{\baselineskip 10pt  \small  Branching fractions of bottomonium $1P \to 1S$ and $2P\to 1S$   Dalitz decays computed using the TFF mass parameters   $a_{b,1P}=0.74$  GeV and $a_{b,2P}=1.41$ GeV.} 
\label{BRbbbarll} 
\end{table} 

\section{Sensitivity of heavy quarkonium  Dalitz modes to a dark photon}\label{darkphoton}
The sensitivity of   Dalitz decays of heavy quarkonia to the contribution of a new light vector mediator,   such as the dark photon ($\gamma^\prime$) or $U$ boson, merits investigations. 
  This mediator  is a prediction  of models extending the  Standard Model with an additional $U(1)^\prime$ gauge symmetry.
 
The proposed dark photon has been connected to anomalies reported by experiments utilising the ATOMKI spectrometer 
\cite{Krasznahorkay:2015iga,Krasznahorkay:2021joi,Krasznahorkay:2022pxs}. Specifically,  an  excess of events was observed in  the distribution of  $e^+ e^-$ pair opening angle at large angles, resulting from the internal conversion  in nuclear de-excitation of  $^8$Be,  $^4$He, $^{12}$C.
The observation was confirmed by the VINATON experiment  \cite{Anh:2024req}. In contrast,  other searches, as the  MEG II experiment, have neither confirmed \cite{MEGII:2024urz} nor excluded the anomaly \cite{Barducci:2025hpg}. Proposed explanations of the ATOMKI anomaly  require a new boson, often designed $X(17)$, with an approximate  mass of $m_X \simeq 17$ MeV.

Excluding other possibilities, the $X(17)$ particle could be the $U(1)^\prime$ gauge mediator with feeble interaction with SM particles \cite{Fabbrichesi:2020wbt,Alves:2023ree}. 
The interaction of this  mediator (the $A^\prime$ field)  with ordinary fermions is described by a  Lagrangian  analogous to the electromagnetic Lagrangian, 
\be  {\cal L}_X= -\sum_f (e \, k_f  \,  J^\mu_f \,A^\prime_\mu) 
\ee 
 where   $J^\mu_f={\bar f } \gamma^\mu f$ is the fermion current.  The dimensionless  parameters $k_f$ govern the coupling strengths and are specified as   
$k_\ell=\epsilon_\ell$ for leptons ($\epsilon_\ell$ is the lepton specific coupling) and  $k_q=e_q \epsilon_q$ for quarks ($e_q$ is the quark electric charge and $\epsilon_q$  the quark specific coupling).   Furthermore,  the new mediator can interact  with the Standard Model electromagnetic field via the kinetic mixing  
\be  
{\cal L}_{mix}=-\frac{\epsilon}{2} F^{\prime}_{ \mu \nu}  F^{\mu \nu}  \,\,,
\ee    
with $\epsilon$  the mixing parameter governing the interaction \cite{Fayet:1980ad,Holdom:1985ag}. 
A small width  is expected for  light $\gamma^\prime$,  with  mass  below $1$ GeV \cite{Bjorken:2009mm,Batell:2009yf,Reece:2009un}. Such a light mediator has been  searched at beam dump experiments \cite{Riordan:1987aw,Bross:1989mp},  fixed target facilities \cite{APEX:2011dww,HADES:2013nab,Merkel:2014avp,NA482:2015wmo,NA62:2019meo,NA62:2023rvm,NA64:2018lsq,NA64:2023wbi},  colliders \cite{OPAL:2002vhf,PHENIX:2014duq,BaBar:2014zli,BESIII:2017fwv,BESIII:2018aao,BESIII:2018qzg,Anastasi:2015qla,KLOE-2:2018kqf,LHCb:2017trq,Belle-II:2022jyy,ATLAS:2022xlo,CMS:2024zqs} including
  PADME experiment at the Frascati INFN National Laboratories \cite{Darme:2022zfw,PADME:2025dla}.  Methods  designed for searching the ATOMKI $X(17)$  signal have been proposed \cite{REDTOP:2022slw,Hostert:2023tkg,Dutta:2024yjp,Gustavino:2024des}. Constraints for the  parameter space have been derived, with the conclusion that the results of the ATOMKI experiment are compatible with the  $e^+ e^-$ coupling $\epsilon_e$  in the range  $\epsilon_e \in[0.2,\,1.4]\times 10^{-3}$ \cite{Feng:2016jff}.
 
Charm decays  are  important processes where to investigate  $X(17) (\gamma^\prime)$   \cite{Li:2009wz,Fu:2011yy,Ban:2020uii,Castro:2021gdf,Lee:2025lwv,Tran:2025fhb}. 
If  produced  in processes such as $M^\prime \to M \gamma^\prime$ followed by $\gamma^\prime \to \ell^+ \ell^-$,  the $\gamma^\prime$ contribution would  modify  the Dalitz decay width  and the dilepton invariant mass distribution. The largest effect would be close to $m_{\gamma^\prime}$ if the mass is in the $q^2$ kinematical range. 

\begin{figure}[t]
\begin{center}
\includegraphics[width = 0.44\textwidth]{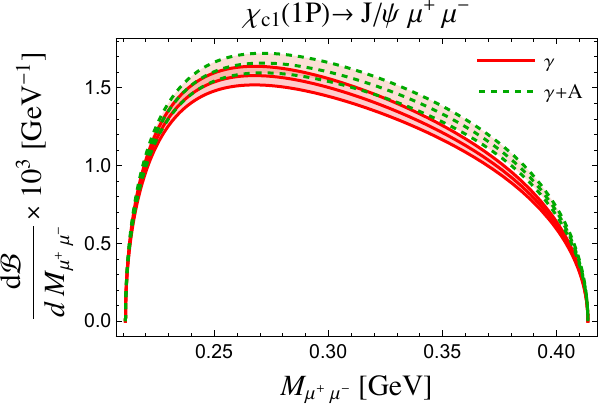} \hskip 0.5cm
\includegraphics[width = 0.44\textwidth]{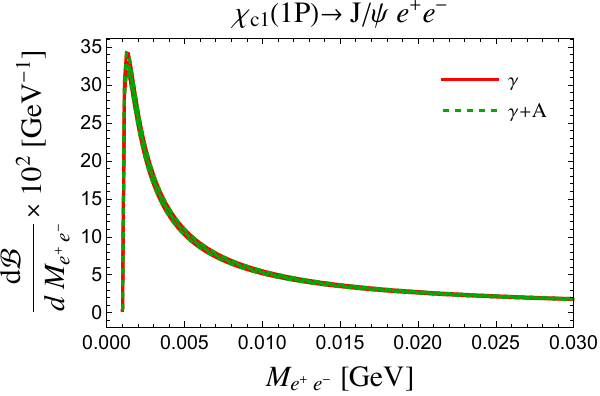} \\ \vskip 0.5cm
\includegraphics[width = 0.44\textwidth]{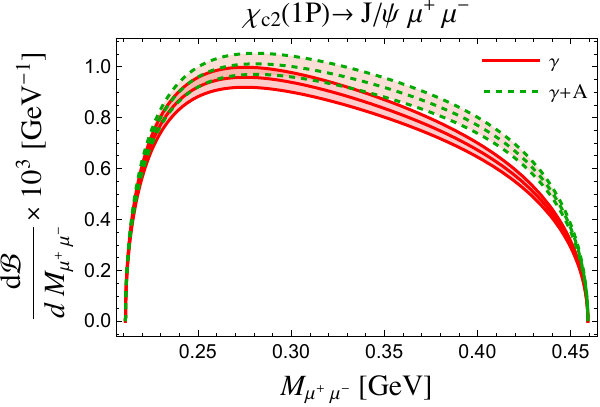} \hskip 0.5cm 
\includegraphics[width = 0.44\textwidth]{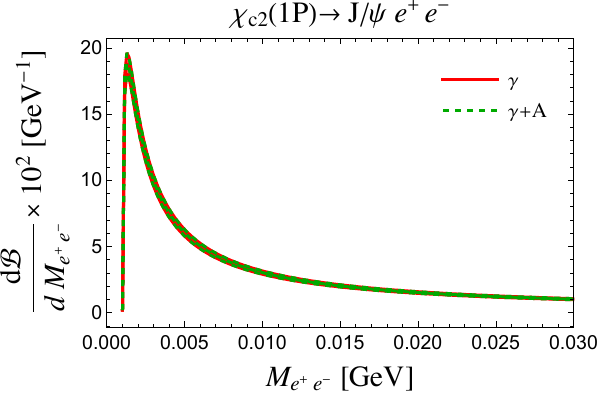}  \\ \vskip 0.5cm
\caption{\baselineskip 10pt  \small Dilepton distributions with (red continuous lines) and without (green dashed lines)  the dark photon contribution in  
$\chi_{c1}(1P)\to J/\psi \ell^+ \ell^-$ (top panels) and $\chi_{c2}(1P)\to J/\psi \ell^+ \ell^-$ (bottom panels). The left plots refer to the muon channel, the right ones to the electron channel.}\label{spettri}
\end{center}
\end{figure}

Experimental studies have been carried out on the channels  $J/\psi \to \eta^{(\prime)} \, \gamma^\prime (e^+ e^-)$  \cite{BESIII:2018aao,BESIII:2018qzg}  and $\chi_{cJ}\to J/\psi \, \gamma^\prime (e^+ e^-)$   (with $J=0,1,2$) \cite{BESIII:2025otp}. 
 Here we analyze the  role of $X(17)(\gamma^\prime)$ in the Dalitz decays  $\chi_{c1,2}(1P) \to J/\psi \ell^+ \ell^-$, with $\ell=\mu,\,e$,  for which  the most precise data are available. 
The  contribution of a new vector mediator of mass $m_X$ and width $\Gamma_X$ modifies Eq.~\eqref{fQED},  
\be
F_{\rm Dark}(q^2)=F_{QED}(q^2) \times \Big | 1+e^{i\phi} \epsilon_c \, \epsilon_e \frac{q^2}{q^2 -m_X^2+ i m_X \Gamma_X} \Big |^2 \,\,, 
\label{fDark}
\ee
with $\phi$ a phase  between the photon and  dark photon amplitudes.
We fit  ${\cal B}(\chi_{c1}(1P) \to J/\psi \, e^+ e^-)$ and ${\cal B}(\chi_{c1}(1P) \to J/\psi \,\mu^+ \mu^-)$  to determine  the parameters in Eq.~\eqref{fDark}  together with TFF mass parameter. The results are used to compute the  $\chi_{c2}$ decay rate.
 We  set $m_X=17$ MeV and $\epsilon_e=1.4 \times 10^{-3}$ at the upper edge of the range obtained in \cite{Feng:2016jff}.  For $\Gamma_X$ we choose the experimental resolution  in the variable $M_{e^+ e^-}$ quoted in  \cite{BESIII:2025otp}: $\Gamma_X \sim 2 \, {\rm MeV}$.   We vary $\epsilon_c$  in a range 
up to $10^{-2}$, considering the upper bound  $\epsilon_c< 1.2 \times 10^{-2}$ at 90\% C.L.  obtained in  \cite{BESIII:2025otp}. The experimental branching fractions  require  $\phi=\pi$. The best fit produces a set of pairs $(a,\, \epsilon_c)$, with  a benchmark point 
 \be
 a=0.71 \,\, {\rm GeV} \,\, , \hskip 1 cm \epsilon_c=3.2 \times 10^{-3}\,\,.
 \label{darkfit}
 \ee
 The inclusion of the dark photon contribution  modifies the preferred value of the mass parameter $a$ in the transition form factor.
 After having verified that at the benchmark point   the Dalitz branching fractions to electrons and muons of the partner state  $\chi_{c2}$ are obtained, we analyse the dilepton   distributions  for $\ell=e,\mu$: they are  displayed in Fig.~\ref{spettri}.
 The impact  on the  dilepton distribution is small, even though non-negligible. A similar effect is found in the $\chi_{c2} \to J/\psi \ell^+ \ell ^-$  distributions, shown   in the same figure.
These results are an indication of the kind of precision requested  to  display  a dark photon effect. Since the deviations are related in all modes,  the sensitivity to the  new contribution is enhanced if the decays are analysed altogether. 

 It would be tempting to  attribute the excess of events observed at low values of $M_{e^+ e^-}$ for $\chi_{c1}$ and $\chi_{c2}$,  as displayed in Fig.~\ref{spectraee}, to a non-Standard Model contribution. Arguments disfavoring this interpretation include the lack of  the related deviation  in the case of $\chi_{c0}$,  the broad shape of the signal and that
 a  deviation of this size would necessitate a very large value of  the coupling $\epsilon_c$. 
The full Monte Carlo simulation of the Dalitz processes described in \cite{BESIII:2025otp} does not report deviations with respect to the measurements. These arguments  induce us to refrain considering the excess as significant.  Nevertheless, the warning message should be kept, these modes warrant continued attention.

\section{Conclusions}\label{conclusions}
We have analysed a set of charmonium and bottomonium Dalitz processes, organised in classes where the decaying mesons belong to the same spin multiplet and also the produced 
states are comprised in the same spin multiplet. This organisation has allowed us to exploit the heavy quark spin symmetry, establishing relations among different modes. The connection with the radiative processes has been systematically used, while the transition form factors have been determined using measurements.  A remarkable overall agreement is found  between the computed branching fractions and dilepton mass distributions with existing measurements. Predictions have been provided for many unobserved modes. The predicted rates are within the reach of the present experimental facilities. We have also investigated the sensitivity of two charmonium Dalitz processes to a dark photon contribution.

\section{Note added}\label{note}
Since this manuscript was posted on  arXiv and submitted for publication, the decay  $\chi_{c1}(3872) \to J/\psi \mu^+ \mu^-$ has been observed by the LHCb Collaboration \cite{LHCb:2026plk}. The measured branching fraction is consistent with the value reported in this study. 

\section*{Acknowledgements}
We thank G.~Roselli for collaboration in the analysis of radiative heavy quarkonium decays  in \cite{Colangelo:2025uhs}. We are grateful to  I.~Belyaev,  M.~Buonsante, M.~Needham  and M.~Pappagallo  for discussions.
This work has  been carried out within the INFN project (Iniziativa Specifica)  SPIF.

\appendix
\numberwithin{equation}{section}
\section{Radiative decay widths of heavy quarkonium}\label{appA}
We collect the expressions of the widths of  electric dipole  decays described by the effective Lagrangian \eqref{lagPS} \cite{Colangelo:2025uhs}.  For a quarkonium state the spectroscopic notation $n^{2s+1}L_J$  is adopted, with $n$ the radial number, $L$  the orbital angular momentum,  $s$ the spin of the quark pair and   $J$ the total spin:
\bea
\Gamma(n^3P_J \to m^3S_1 \, \gamma) &=& {(\delta_Q^{nPmS})^2 \over 3 \pi} k_\gamma^3{M_{S_1} \over M_{P_J}}   \label{deltanPmS1} \\
\Gamma(n^1P_1 \to m^1S_0 \, \gamma) &=& {(\delta_Q^{nPmS})^2 \over 3 \pi} k_\gamma^3{M_{S_0} \over M_{P_1}} \label{deltanPmS3}\\
\Gamma(m^3S_1 \to n^3P_J \, \gamma) &=& (2J+1) {(\delta_Q^{nPmS})^2 \over 9 \pi} k_\gamma^3{M_{P_J} \over M_{S_1}}  \label{deltanPmS2} \\
\Gamma(m^1S_0 \to n^1P_1 \, \gamma) &=& {(\delta_Q^{nPmS})^2 \over \pi} k_\gamma^3{M_{P_1} \over M_{S_0}}  \,\,\ . \label{deltanPmS4}
 \eea 
$M_i$ are the quarkonium masses and  $k_\gamma$ is the photon energy.  Due to the heavy quark spin symmetry, the same coupling $\delta_Q^{nPmS}$ governs all channels.
The width of the  magnetic dipole decay  described by the effective Lagrangian \eqref{lagSS} reads 
\be
\Gamma(n^3S_1 \to m^1S_0 \gamma) = {4(\delta_Q^{nSmS})^2 \over 3 \pi} k_\gamma^3{M_{S_0} \over M_{S_1}}   \,\,.\label{deltanSmS} \\
\ee
\bibliographystyle{JHEP}
\bibliography{refFPR}
\end{document}